\def\rN{\rho_N}

\def\al{\alpha_c}

\def\bunit{{\rm MeV\,fm^{-3}}}
\def\bmax{B_{\rm max}}

\def\drom{{\rm d}}

\def\msol{M_{\odot}}

\def\rhob{\rho_{\rm b}}
\def\Pb{P_{\rm b}}
\def\pnd{P_{\rm ND}}
\def\rond{\rho_{\rm ND}}

\def\al{\alpha_c}

\def\rN{\rho_N}

\def\bunit{{\rm MeV\,fm^{-3}}}
\def\rounit{{\rm g\,cm^{-3}}}

\def\rcore{R_{\rm core}}
\def\mcore{M_{\rm core}}
\def\rmin{R_{\rm min}}
\def\rhb{\overline{\rho}_0}
\def\chio{\chi}
\def\rap{R_{\infty}}

\documentclass{aa}
\usepackage{graphics}
\usepackage{natbib}
\bibpunct{(}{)}{;}{a}{}{,}

\input{epsf}
\begin{document}

\title{On the minimum radius of strange stars with crust}
\author{J.L. Zdunik}

\institute{N. Copernicus Astronomical Center, Polish
           Academy of Sciences, Bartycka 18, PL-00-716 Warszawa,
           Poland\\ e-mail:jlz@camk.edu.pl}
%
%
\date{Received 29 May 2002 / Accepted 9 August 2002}
%

\abstract{
  The minimum value of the radius of strange star covered
by the crust of nuclear matter is determined. The results for the maximum
 possible thickness of the crust (up to the
neutron drip) as well as the possibility of thinner crust postulated by some authors
are discussed.
The minimum radius of the strange star with maximal crust is 5.5 km. The useful scaling
formulae with respect to the main parameters describing strange matter
and the density at the bottom of the crust are presented.
\keywords{dense matter -- equation of state  -- stars: neutron}
}
\maketitle

\section{Introduction}

The idea of the compact stars build of strange matter was
presented by \cite{witten84} and the models of
stars were calculated using various models of strange
matter by \cite{hzs86} and
\cite{afo86}.
The main idea is that the u,d,s matter is ground state of matter
at zero pressure (self-bound strange quark
 matter) i.e.  its energy per baryon is smaller than that of iron:
\begin{equation}
\mu_0\equiv\mu(P=0) < {\rm M}({\rm ^{56}Fe})\,c^2/56=930.4~{\rm MeV}
\label{udsstab}
\end{equation}
where ${\rm M}({\rm ^{56}Fe})$ is the mass of the ${\rm ^{56}Fe}$ atom.

Recently the increasing interest in strange stars is connected
with some estimations of the radius of the isolated neutron
(or strange) stars from some limits on the temperature of such a star.
Some indications that the radius is small
favor strange stars as a possible explanation.

Bare strange star surface is very poor emitter of photons with
energies lower than $\sim 20$~MeV \citep{afo86,chs91}.
However strange star could be covered by a crust of nuclear matter,
which changes the properties of the star allowing for the emission of photons of
lower energies and the black body radiation from the stellar surface \citep{afo86}.
The additional mechanism of the radiation from bare strange
star due to the $e^+e^-$ pair creation has recently been proposed
by \cite{usov98,usov01}, but its contribution for $T<10^9$~K is
negligible.

As a result of observational data one can try to determine the
apparent radius of the neutron star $\rap$. If $\rap$ would
be smaller than $\sim 12$~km the only solution seems to be the strange star
\citep{haensel01}.
Recent observations of the isolated neutron star candidate RX
J1856.5-3754 have been interpreted as a star with the radius
$\rap \sim 3.8-8.2$~km \citep{drake02} (but see also \cite{walter02}).

In this paper I present some limits on the radius of strange stars
if they are covered by the crust of nuclear matter.

\section{Strange stars with crust}

\subsection{Equations of state}

In this paper I
consider two types of EOSs of strange
matter: MIT Bag Model and models presented by \cite{dey98}.

In the framework of the phenomenological MIT bag
model the quark matter is the
mixture of the massless $u$ and $d$ quarks, electrons and massive
$s$ quarks. The model is described in detail in  \cite{fj84},
where the formulae for physical parameters of a strange matter
are also presented.
There are the following physical quantities entering this model:
$B$ -- the bag constant, $\al$ -- the QCD coupling constant and $m_s$ --
the mass of the strange quark. It is necessary to introduce also
the parameter $\rN$ -- the renormalization  point. Following \cite{fj84}
 we choose $\rN=313$~MeV.

The consequence of this model of strange matter is scaling of all
thermodynamic functions and parameters of the strange stars
(mass, radius etc.) with some powers of $B$ \citep{hzs86,zdunik00}.

The main model considered in this paper corresponds to the following set of the
MIT Bag Model model parameters for strange matter:  bag constant $B=56~{\rm
MeV/fm^3}$, mass of the strange quark $m_{\rm s}=200~{\rm MeV/c^2}$,
and QCD coupling constant $\alpha_{\rm c}=0.2$.
This EOS of strange quark matter (called SQM1 as in \cite{ZHG01})
had been also used in \cite{zdunik00} and \cite{ZG01}).
It yields an energy per unit baryon
number at zero pressure $E_0=918.8 ~{\rm MeV} <E(^{56}{\rm
Fe})=930.4~$MeV. The maximum allowable mass for
strange stars is $M_{\rm max}^{\rm stat}=1.8~{\rm M_\odot}$.

I also consider two cases of strange matter based on MIT Bag Models
for massless strange
quarks in which the dependence $P(\rho)$ is exactly linear and
scaling laws with $B$ are exact \citep{zdunik00}. These two models
correspond to "standard" value of the bag constant $B=60~\bunit$
and to the maximum possible value of $B$ consistent with the
requirement of the stability of strange matter [Eq. (\ref{udsstab})]:
$\bmax=91.5~\bunit$.

The second EOS of strange matter considered in this paper is
based on the model of strange matter proposed by
\cite{dey98} which incorporates restoration of
chiral quark masses at high densities. Two sets of parameters
describing this model have been used to determine properties of
strange stars \citep[see eg.][]{li99a,li99b} and usually referred
to as SS1 and SS2.

The main property of the self bound strange matter is the large
density at zero pressure and the equation of state which can be
very accurately approximated by the linear dependence $P(\rho)$.

As it has been shown in \cite{zdunik00,gondek00} in both cases
(MIT and \cite{dey98} models) with accuracy of the order of 1-2 \%
we can write:
\begin{equation}
P=a c^2 (\rho-\rho_0)
\label{eossqm}
\end{equation}
where $a\simeq 1/3$ and $\rho_0$ is the density of self bound
strange matter at zero pressure. Both $a$ and $\rho_0$ are
functions of the parameters describing the model of strange matter.

In all considered models the crust is described by the BPS model
of dense matter below neutron drip \cite{BPS}. The maximum
pressure and density at the bottom of the crust which can be
formed on strange star are defined by the neutron drip point
\citep{afo86} and are equal to:
$\pnd=7.8\,10^{29} {\rm dyn\, cm^{-2}}$,
$\rond= 4.3\,10^{11} {\rm g\, cm^{-3}}$.

As it has been pointed out by \cite{afo86} the width of the
gap between the strange core and the crust could determine the density
at the bottom of the crust and the crust could be significantly
thinner \citep[see e.g.][]{hl97,phukon00}.
We can study this effect by considering different values of the pressure
$\Pb$
(and density $\rhob$) at the bottom of the crust i.e.
treating $P_{\rm b}$ as a free parameter
resulting from other considerations, not necessarily equal to $\pnd$.
Of course the condition $\Pb < \pnd$ has to be fulfilled.

\subsection{The role of the size of the crust}

The strange star configurations are calculated by solving
Oppenheimer-Volkoff equations in the case of spherical symmetry.

\begin{equation}
{\drom P\over \drom r}=-{Gm\rho\over r^2 }\left(1-{2Gm\over rc^2}\right)^{-1}
\left(1+{P\over \rho c^2}\right)\left(1+{4 \pi r^3 P\over m c^2}\right)
\label{tov}
\end{equation}

It has been shown by \cite{ZHG01} that for strange stars
the thickness of the crust can be well described by the approximate
formula valid for large and intermediate values of the stellar mass.

The formula presented in \cite{ZHG01} does not render the properties
of the $M(R)$ dependence for small stellar masses (say $ M < 0.5 \msol$),
especially the existence of the minimum radius of the star.

This effect could be quite easily obtained by more careful
approximation of the Eq. (\ref{tov}) in the crust region.
The essential point is that although
for small masses (close to the minimum radius) the mass is concentrated
in the strange core (more than 99.9 \%) the thickness of the crust is relatively
large compared to the core radius. Thus in approximation we can assume that
$M={\rm const}=\mcore$ but we have to take into account the changes of
$r$ thorough the crust. This effect
has not been
considered
in our previous paper \citep{ZHG01}.
We can safely neglect in the crust the last two
terms of Eq. (\ref{tov}). In the crust the maximum values of the factors $P/\rho c^2$ and
${4 \pi r^3 P/ m c^2}$ are of the orders $10^{-3}$ and  $10^{-6}$ respectively.
As a result we obtain:

\begin{figure}     %
\resizebox{\hsize}{!}
{{\includegraphics{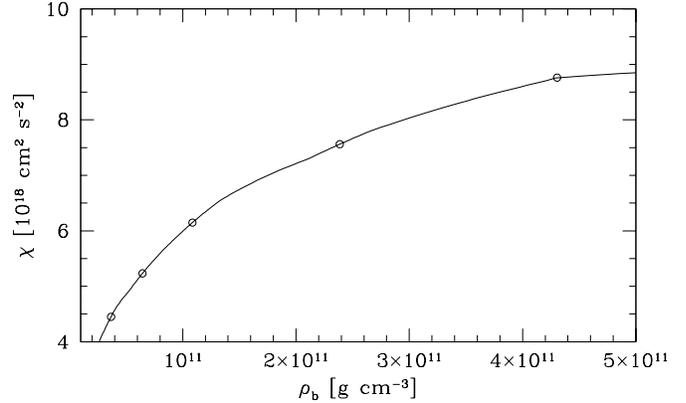}}}
\caption[]{The function $\chio=\int_0^{P_{\rm b}} \rho^{-1} d P$ for
the BPS equation of state of the outer crust. The open circles denote the
parameters of the bottom of the crust
for which the structure of the stellar configurations has been
calculated (see Fig. \ref{mrsqm1} and table \ref{tabsqm1}).
}
\label{figchi}
\end{figure}

\begin{equation}
{\drom P\over \rho}=-GM {\drom r \over r^2 (1-{2GM\over rc^2})}
\label{tovcr}
\end{equation}
which results in the following equation for the stellar radius $R$:

\begin{equation}
\chio={1\over2}\,c^2\,\ln\left[{1-{2GM\over Rc^2}\over
1-{2GM\over \rcore c^2}}\right]
\label{rap}
\end{equation}
where
\begin{equation}
\chio=\int_0^{P_{\rm b}} {\drom P\over \rho}
\label{chio}
\end{equation}
is thermodynamic potential characteristic to the equation of state
and $P_{\rm b}$ is the pressure at the bottom of the crust.

The function $\chio(\rho_{\rm b})$ is plotted in Fig. \ref{figchi}.

\begin{figure}     %
\resizebox{\hsize}{!}
{{\includegraphics{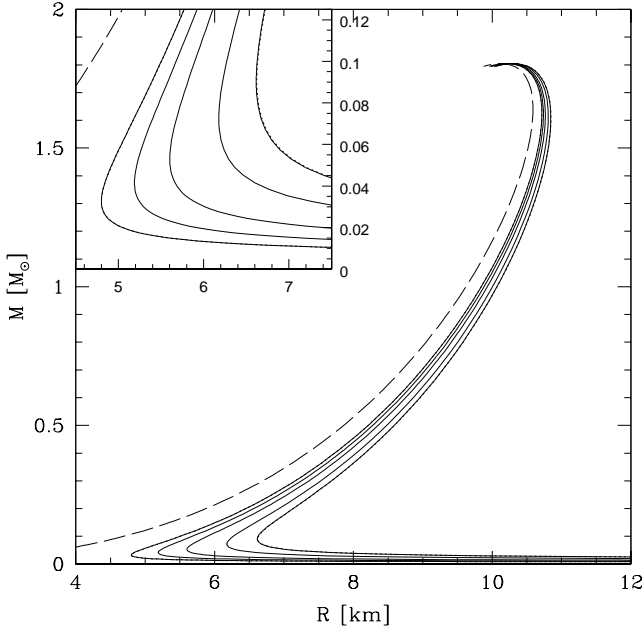}}}
\caption[]{The mass vs radius relation for strange stars with crust for different
choices of the pressure at the bottom of the crust. From the right to the
left $\Pb$ is equal to 100\%, 50\%, 20\%, 10\% and 5\%
of the pressure at the neutron drip
point $\pnd=7.8\,10^{29} {\rm dyn\, cm^{-2}}$.
The corresponding densities are given in Table \ref{tabsqm1}.
The dashed line corresponds to the bare strange star.
The dotted lines presenting the results of the approximate Eq.
(\ref{rapsol}) for the cases 100\% and 5\% are nearly
undistinguishable from the exact solutions (solid lines)
(the small difference could be seen for the $\Pb=\pnd$ case in
the insert showing the enlarged region of the minimum radius).
The EOS of strange matter is given by the model SQM1 (see the text).
}
\label{mrsqm1}
\end{figure}

The solution of  Eq. (\ref{rap}) can be written in the form:
\begin{equation}
{1\over R}={1\over \rcore}\exp{(2\chio/c^2)}+{c^2\over 2GM}
(1-\exp{(2\chio/c^2)})
\label{rapsol}
\end{equation}
which in the limit up to the first order in $1/c^2$
results in the formula:
\begin{equation}
{1\over R}={1\over \rcore}-{\chio\over GM}+
2{\chio\over c^2}\left({1\over\rcore}-{\chio\over 2GM}\right)
\label{rapsolpn}
\end{equation}

In Fig. \ref{mrsqm1} one can see the accuracy of our approximation
\ref{rapsol} in which we have used the values of the mass and radius of bare
strange star ($\rcore$ and $\mcore$ depicted by the dashed line)
to determine the radius of
strange star with crust (the mass is assumed to have the same
value). In Fig. \ref{mrsqm1} the results of Eq. (\ref{rapsol}) cannot be distinguished
from the exact solution (the very small difference can be seen in
the insert showing the  enlarged region near to the minimum radius).

The minimum radius of the star can be easily obtained by
differentiating Eq. (\ref{rapsol}) with respect to $M$
assuming that $M\sim \rcore^3$ which for self bound stars of small
masses is a very good approximation.
As a result we obtain:
\begin{eqnarray}
R&=&{3\over2}\rcore\exp{(-2\chio/c^2)}\\
{GM\over Rc^2}&=&\exp{(2\chio/c^2)}-1
\label{rapsol1}
\end{eqnarray}

If we are interested in the minimum radius of the strange star with crust we
can safely work in the nonrelativistic limit (the relativistic parameter
$GM/Rc^2$ is of the order of 2\%).
In nonrelativistic limit we obtain:
\begin{eqnarray}
{GM\over \rcore}&=&{4\over 3} \pi G \rho_0 \rcore^2=3\chio\\
R&=&{3\over2}\rcore
\label{rapsolnr}
\end{eqnarray}

It should be noted that the determination of
the minimum radius in nonrelativistic limit is consistent with the
assumption of the constant density of the core build of strange matter.
For the small mass strange stars with EOS \ref{eossqm}
the density in the center is larger then
$\rho_0$ by $\delta \rho$ given by the formula:
\begin{equation}
{\delta\rho\over \rho_0}\simeq{1\over2 a}{G\mcore\over \rcore c^2}
\label{delrho}
\end{equation}
which follows from the expansion of the equation of hydrostatic
equilibrium in $\delta\rho/\rho$ in Newtonian limit.
The central pressure is given by:
\begin{equation}
P_c\simeq{1\over2 }{G\mcore\rho_0\over \rcore}
\label{pcap}
\end{equation}
In this approximation the mass of the strange core is:
\begin{eqnarray}
\mcore&=&M_0\left(1+{1\over5 a}{GM_0\over  \rcore c^2}\right)\\
M_0&=&{4\over 3}\pi\rho_0\rcore^3
\label{mcore}
\end{eqnarray}

The interesting consequence of Eq. (\ref{delrho}) is 
that the self-bound star with crust   reaches its minimum radius at
the central density $\rho_c\simeq\rho_0\,(1+1.5\chio/ac^2)$ and central
pressure $P_c\simeq 1.5\chio\rho_0$. The accuracy of this
expansion at this mass is 10\% for $\delta\rho/\rho_0$
which corresponds to 10\% error in $P_c$ and 1\% in $\rho_c$.

In Fig. \ref{mrsqm1} we present mass versus radius relations for the SQM1 Eos of the quark
matter and different choices of $P_{\rm b}$ equal to 1, 0.5, 0.2, 0.1, 0.05 of the pressure
at the neutron drip point.

\begin{table*}
$$
\begin{array}{@{\extracolsep{5pt}}cccl*6c}
\hline
&&&\vline&&&&&&\\
\omit\vspace{-8truept}\\
P_{\rm b}/c^2& \rho_{\rm b}& \chio&\vline& \rho_c&  R& M& z_s& \rcore/R& \mcore/M\\
\rounit&\rounit&{\rm cm^2\,s^{-2}}&\vline&\rounit&{\rm km}&\msol &&&\\
\hline
&&&\vline&&&&&&\\
\omit\vspace{-8truept}\\
8.6787\,10^{8}& 4.3051\,10^{11}& 8.761\,10^{18}&\vline& 4.751\,10^{14}&6.616&0.09216& 0.0208&0.6856&0.999765\\
4.3393\,10^{8}& 2.3864\,10^{11}& 7.561\,10^{18}&\vline& 4.712\,10^{14}&6.177&0.07348& 0.0177&0.6822&0.999860\\
1.7357\,10^{8}& 1.0856\,10^{11}& 6.148\,10^{18}&\vline& 4.670\,10^{14}&5.602&0.05363& 0.0143&0.6791&0.999929\\
0.8679\,10^{8}& 0.6452\,10^{11}& 5.233\,10^{18}&\vline& 4.643\,10^{14}&5.188&0.04229& 0.0121&0.6778&0.999958\\
0.4339\,10^{8}& 0.3676\,10^{11}& 4.450\,10^{18}&\vline& 4.621\,10^{14}&4.800&0.03300& 0.0102&0.6759&0.999975\\
\hline
\end{array}
$$
\caption[]{The parameters of the strange star at the minimum
radius for different choices of the pressure at the bottom of the crust
(first column). The first three columns characterize the properties of the crust -
the pressure and density at the bottom of the crust.
The model of strange matter is SQM1 (see text) and the crust is described by the BPS
equation of state.}
\label{tabsqm1}
\end{table*}

The parameters of stellar configurations at the point with minimum
radius are presented in Table \ref{tabsqm1}. The gravitational
redshift of photon emitted from from the surface
$$z_s=(1-2GM/Rc^2)^{-1/2}-1$$
measures the importance of relativistic effects. As we see this value is
comparable to the departure of $\rcore/R$ from the Newtonian value
$2/3$ [Eq. (\ref{rapsolnr})]. The factor $(1+z_s)$ connects also the
radius of the star and the "apparent radius":
\begin{equation}
\rap=R\,(1+z_s)
\label{rappar}
\end{equation}

At its minimum value the "apparent radius" is larger than the
radius of the star by $\sim 1-2 \%$.

\begin{figure}     %
\resizebox{\hsize}{!}
{{\includegraphics{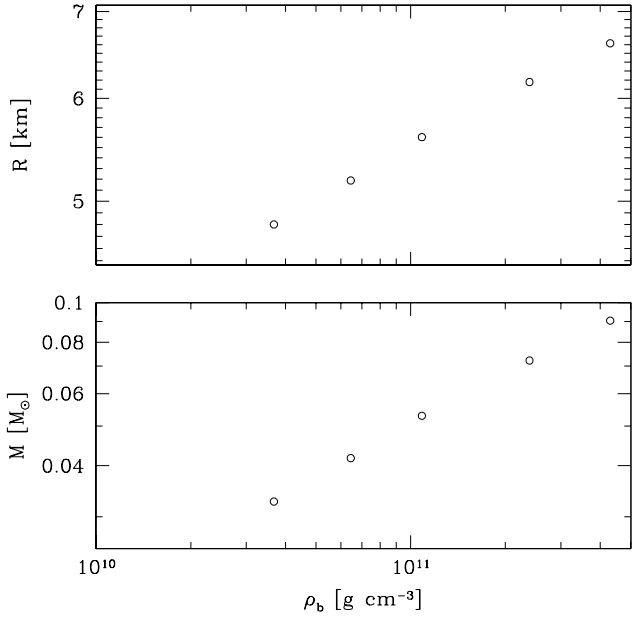}}}
\caption[]{The minimum radius of the star (top) and corresponding mass (bottom)
as a function of the density at the bottom of the crust.
}
\label{rmrho0}
\end{figure}

The dependence of $\rmin$ and $M(\rmin)$ on $P_{\rm b}$ and $\rho_{\rm b}$  is nearly
power-law (see Fig. \ref{rmrho0}),
due to the fact the equation of state of the crust
can be very well approximated by the polytrope.
For polytropic EOS we have:
\begin{equation}
P\sim \rho^\gamma~~~~~~~~\chio={\gamma \over \gamma-1}{P\over \rho}
\label{chiopoly}
\end{equation}
From Eq. (\ref{rapsolnr}) we see that for polytropic EOS
 $\rmin\sim {P_{\rm b}}^{(\gamma-1)/2\gamma}\sim {\rho_{\rm b}}^{(\gamma-1)/2}$ and $M(\rmin)\sim
{P_{\rm b}}^{3(\gamma-1)/2\gamma}\sim {\rho_{\rm b}}^{ 3(\gamma-1)/2}$.

\subsection{The role of the strange matter EOS}
\begin{figure}     %
\resizebox{\hsize}{!}
{{\includegraphics{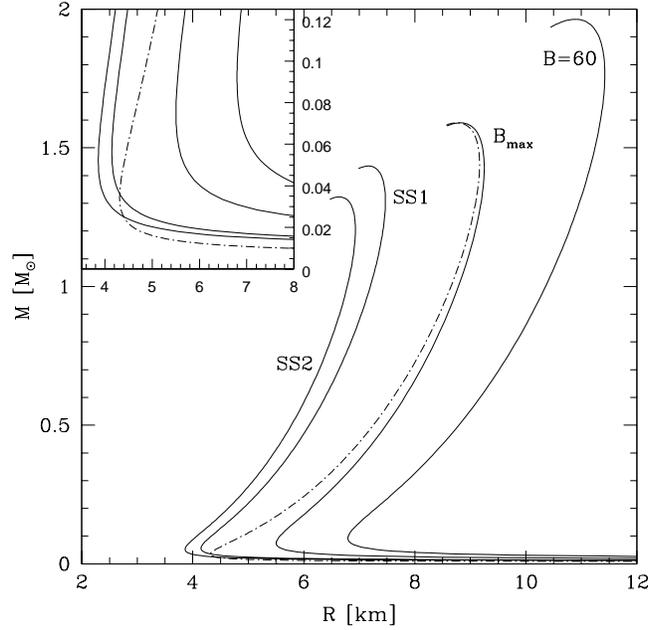}}}
\caption[]{The mass vs radius relation for strange stars with maximum crust for
two choices of the bag constant in the MIT model of strange matter
and for two models of \cite{dey98}. The dot-dashed line
corresponds to the pressure at the bottom of the crust equal to
$0.1\pnd$}
\label{mr0}
\end{figure}

The mass vs radius relation for different models of strange matter
is presented in Fig. \ref{mr0} and the main parameters of these
models ($\rho_0$, $a$) and stellar configurations at minimum radius
are given in table \ref{tabsqm0}

\begin{table*}
$$
\begin{array}{@{\extracolsep{5pt}}cccl*6c}
\hline
&&&\vline&&&&&&\\
\omit\vspace{-8truept}\\
\hbox{Model}&a& \rho_0& \vline& \rho_c&  R& M& z_s& \rcore/R& \mcore/M\\
&&\rounit&\vline&\rounit&{\rm km}&\msol &&&\\
\hline
&&&\vline&&&&&&\\
\omit\vspace{-8truept}\\
B=60.0&1/3  &4.278\,10^{14}&\vline& 4.486\,10^{14}& 6.795&0.09235&0.02070& 0.6844&0.999750\\
B=91.5&1/3  &6.523\,10^{14}&\vline& 6.839\,10^{14}& 5.503&0.07452&0.02062& 0.6836&0.999836\\
\hbox{SQM1}&0.301&4.500\,10^{14}&\vline& 4.751\,10^{14}& 6.616&0.09041&0.02081& 0.6856&0.999765\\
\hbox{SS1}&0.463&1.154\,10^{15}&\vline& 1.194\,10^{15}& 4.148&0.05607&0.02058& 0.6831&0.999906\\
\hbox{SS2}&0.455&1.332\,10^{15}&\vline& 1.378\,10^{15}& 3.861&0.05221&0.02059& 0.6832&0.999919\\
\hline
\end{array}
$$
\caption[]{The parameters of the strange star at the minimum
radius for different choices of the parameters of strange matter.
The first two columns characterize the EOS of the strange star core ($P=a c^2 (\rho-\rho_0)$).
The bottom of the crust corresponds to the neutron drip point
(the maximum crust) and the parameter of the crust ($P_{\rm b}$, $\rho_{\rm b}$, $\chio$)
are given in the first row of Table 1.}
\label{tabsqm0}
\end{table*}

The radius of the strange star with crust at its minimum point is
a very simple functional of the equation of state of strange matter.
In the nonrelativistic limit it depends on the value of the density
of the matter  at zero pressure $\rho_0$ independently of the
specific EOS of the self-bound matter. The parameter $a$
corresponding to the sound velocity of the matter enters next term
in Eq. (\ref{mcore})
via the slight increase of the density in the center of the
strange core.
For given $\chio$ we obtain from Eq. (\ref{rapsolnr}):
\begin{equation}
\rmin={9\over4}\,\sqrt{\chio\over\pi
G}\,\rho_0^{-1/2}=4.9\,\sqrt{{\chio_{}}_{18}\over{\rho_0}_{14}}~{\rm km}
\label{rminnr}
\end{equation}
where ${\chio_{}}_{18}$ and  ${\rho_0}_{14}$ denotes $\chio$ and
$\rho_0$ in units $10^{18}~{\rm cm^2\,s^{-2}}$ and $10^{14}~{\rm
g\,cm^{-3}}$ respectively.

As we see from Eq. (\ref{rminnr}) the minimum radius corresponds to
the maximum $\rho_0$. In the MIT bag model of strange matter
$\rho_0$ is limited by the assumption of the stability of the
matter at zero pressure [Eq. (\ref{udsstab})]. The maximum $\rho_0$
corresponds then to $B=91.5~\bunit$ for massless strange quarks.
The numerical values of the stars at minimum radius as a function
of $\rho_0$ are presented in Table \ref{tabsqm0}.
For other model of the self bound matter the parameters of the
star with minimum radius can be easily obtained from the table
\ref{tabsqm0} using scaling relations with $\rho_0^{-1/2}$
\begin{eqnarray}
\rmin(\rhb)&=&\rmin(\rho_0)\sqrt{\rho_0\over\rhb}\\
M(\rmin(\rhb))&=&M(\rmin(\rho_0))\sqrt{\rho_0\over\rhb}
\label{sclrho0}
\end{eqnarray}

As can be seen from the Table \ref{tabsqm0} the accuracy of these
scalings is better than 1\%.

The surface redshift at minimum radius depends only on the value of
$\chio$
and for fixed density at the bottom
of the crust the points with minimum radius and different $B$ (or
$\rho_0$) lie in the $M(R)$ plane on the straight line
$GM/Rc^2={\rm const}$.
The small differences in $z$ in table \ref{tabsqm0}
reflects the accuracy of our approximations in Newtonian approach.

\section{Conclusions}

The radius of the the strange star with crust can be very accurately calculated
from the parameters of the bare strange stars. Presented formula
[Eq. (\ref{rap})] gives us the stellar radius for a very wide range
of stellar masses from the maximum one down to the masses $\sim
0.02\msol$. The only important assumption is the concentration of
the mass in the core built of strange matter. This assumption is
fulfilled well below the star with minimum radius, thus the
minimum radius of the star can be safely calculated using
presented equations. It should be however mentioned that this
method fails in determination of the point with minimum mass  since
 the mass of crust plays crucial role there.

\begin{acknowledgements}
 This research was partially supported by the KBN grant No. 2P03D.020.20.
I am very grateful to P.~Haensel for careful reading of manuscript and helpful
comments and suggestions.
\end{acknowledgements}

\end{document}